\RequirePackage{fix-cm}
\documentclass[smallextended,referee]{svjour3}       % onecolumn (second format)
\smartqed  % flush right qed marks, e.g. at end of proof
\usepackage{graphicx}
\usepackage{amsmath,amssymb,bm}
\begin{document}

\title{A model of competition among more than two languages%\thanks{Grants or other notes
}


\author{Ryo~Fujie \and Kazuyuki~Aihara \and Naoki~Masuda%etc.
}


\institute{R. Fujie \and K. Aihara \at
              FIRST, Aihara Innovative Mathematical Modelling Project, Japan Science and Technology Agency, 4-6-1 Komaba, Meguro, Tokyo 153-8505, Japan \\
              Tel.: +81-3-5841-6942\\
              \email{fujie@sat.t.u-tokyo.ac.jp}           %  \\
           \and
           R. Fujie \and K. Aihara \at
              Institute of Industrial Science, The University of Tokyo, 4-6-1 Komaba, Meguro, Tokyo 153-8505, Japan
              \and
           K. Aihara \and N. Masuda \at
              Department of Mathematical Informatics, The University of Tokyo, 7-3-1 Hongo, Bunkyo, Tokyo 113-8656, Japan
              }

\date{Received: date / Accepted: date}

\maketitle

\begin{abstract}
  We extend the Abrams--Strogatz model for competition between two
  languages [Nature {\bf 424}, 900 (2003)] to the case of $n (\ge2)$
  competing states (i.e., languages).
 Although the Abrams--Strogatz model for $n=2$ can
  be interpreted as modeling either majority preference or minority
  aversion, the two mechanisms are distinct when $n\ge 3$.  We
  find that the condition for the coexistence of different states is
  independent of $n$ under the pure majority preference, whereas it
  depends on $n$ under the pure minority aversion. We also show that
  the stable coexistence equilibrium and stable monopoly equilibria
  can be multistable under the minority aversion and not under the majority
  preference.  Furthermore, we obtain the phase diagram of the model
  when the effects of the majority preference and minority
  aversion are mixed, under the condition that different states have
  the same attractiveness. We show that the multistability is a generic
  property of the model facilitated by large $n$.
\end{abstract}

\section{Introduction}

The consensus problem, in which we ask whether
the unanimity of
one among different competing states (e.g., opinions) is reached,
and its mechanisms are of
interest in various disciplines including political science,
sociology, and statistical physics. In models of consensus formation,
it is usually assumed that each individual possesses one of the different
states that can flip over time. The flip rate depends on the
environment such as the number of the individuals that adopt a
different state.  Statistical physicists have approached this problem
by analyzing a variety of models including the voter model, majority
rule models, the bounded confidence model, Axelrod's model, and the naming game
(see \cite{sociophys} for a review).

A major mechanism that would lead to consensus in a population is
preference for the majority. 
Collective opinion formation under various majority voting rules has been
examined for mean-field populations
\cite{Chen1,Galam1986,Galam1999,Galam2002,Krapivsky,Lambiotte4,Mobilia} and in different types of
networks such as regular lattices \cite{Chen2,Chen1,deOliveira,Liggett,Tessone}, small-world networks
\cite{PPLi}, heterogeneous networks \cite{Lambiotte3}, and networks
with community structure \cite{Lambiotte1,Lambiotte2}. 
The majority preference may
be identified with the aversion to the minority.  When there are
only two states, they are equivalent
because one state is the majority if and only if the other state is the
minority.  However, the two principles may be distinct when more than
two states are assumed \cite{Volovik2009EPL}.  We are concerned with
this case in the present study.

Language competition is an example of consensus problems.
The model proposed by Abrams and Strogatz (AS model)
accounts for extinction of endangered
languages \cite{Abrams1}. The AS model implements
competition between two
languages for speakers in a population.
The dynamics of the model is based on the majority preference, which is also regarded as the minority aversion because there are just two competing languages.

Several authors found
that different languages can stably coexist in variants of the AS model.
Two languages can coexist by spatial segregation in a model in which
competition dynamics and spatial diffusion are combined \cite{Patriarca1,Patriarca2}. A Lotka--Volterra variant of the AS model also leads to coexistence \cite{Pinasco}. Introduction of bilingual individuals also enables
coexistence \cite{Mira1,Mira2}.
Castell\'o and colleagues investigated
variants of the AS model with bilingualism on various networks
\cite{Castello1,Castello2,Castello3,Castello4}. 
Coexistence also occurs when
the attractiveness of languages is dynamically manipulated
\cite{Chapel} or when 
bilingualism, horizontal (i.e., from adults to adults) and vertical
(i.e., from adults to children) transmission of languages, and
dynamics of the languages' attractiveness are combined
\cite{Minett,Wang}.

In the present work, we extend the AS model to the case of competition
among a general number of languages, denoted by $n$.
Our model is a mean-field model (i.e., without spatial or network structure), as is the original AS model.
Because the AS model has been used for modeling competition of other cultural
or social traits such as religion \cite{Abrams2},
opinion, and service sectors \cite{Li},
we use the term ``state'' instead of ``language'' in the
following. We show that the behavior of the
model is essentially different between $n=2$ and $n\ge 3$.
In particular, the coexistence of different states and the consensus can be
multistable, i.e., the coexistence and consensus equilibria are both stable, only when $n\ge 3$.

\section{Model}
\label{model}

We extend the AS model to the case of competition among $n (\ge2)$ states. The dynamics of the fraction of state $i$ ($1\le i\le n$) is given by
\begin{equation}
\frac{dx_i}{dt}=\sum_{j=1, j\neq i}^n x_jP_{ji}-x_i\sum_{j=1, j\neq i}^n P_{ij},
\label{dx/dt}
\end{equation}
where $x_i$ is the fraction of state $i$ in the population,
and $P_{ji}$ represents the transition rate from state $j$ to state $i$. Equation~\eqref{dx/dt} respects the conservation law $\sum_{i=1}^n x_i=1$. The transition rates of the original AS model (i.e., $n=2$) are given by
\begin{equation}
P_{ji}=cs_ix_i^a =cs_i(1-x_j)^a\quad \left((i,j)=(1,2), (2,1)\right),
\label{originalP}
\end{equation}
where $a>0$ controls the strength of the frequency-dependent state transition, $s_i>0$ is the attractiveness of state $i$, and $\sum_{i=1}^n s_i=1$ \cite{Abrams1}.
Because $c$ simply specifies the time scale of the dynamics,
we set $c=1$.

Equation~\eqref{originalP} allows two interpretations:
majority preference because
$P_{ji}=s_ix_i^a$ and minority aversion because
$P_{ji}=s_i(1-x_j)^a$.
The two principles lead to the same model when $n=2$.
However, the two principles are distinct when $n\ge 3$.
Therefore, we redefine $P_{ji}$ to allow for independent manipulation of
the two 
factors. The transition rates of the extended model are defined by
\begin{equation}
P_{ji}=s_ix_i^\beta(1-x_j)^{a-\beta},
\label{P}
\end{equation}
where $\beta (\ge 0)$ and $a-\beta (\ge 0)$ represent the strength of the majority preference and the minority aversion, respectively.  When $n=2$, the dynamics given by the substitution of
Eq.~\eqref{P} in Eq.~\eqref{dx/dt} becomes independent of the
$\beta$ value.

\section{Analysis}
\label{analysis}

\subsection{Case of majority preference (i.e., $\beta=a$)}\label{sub:majority preference}

In this section, we set
$\beta=a$ to analyze the case in which the
majority preference is present and the minority aversion is absent.
Substitution of Eq.~\eqref{P} in Eq.~\eqref{dx/dt} yields
\begin{eqnarray}
\frac{dx_i}{dt}&=&\sum_{j=1, j\neq i}^n x_js_ix_i^a - x_i\sum_{j=1, j\neq i}^n s_jx_j^a \nonumber\\
&=&\left(s_ix_i^{a-1}-\left<sx^{a-1}\right>\right)x_i,
\label{cp4m}
\end{eqnarray}
where $\left< \cdot \right>$ represents the average over the
population, i.e., the average of a state-dependent variable with weight $x_i$ ($1\le i\le n$). 
Equation~\eqref{cp4m} is a replicator equation \cite{Hofbauer1998}
in which $s_ix_i^{a-1}$ and $\left<sx^{a-1}\right>=\sum_{\ell=1}^n s_{\ell}x_{\ell}^a$ play the role of the fitness for state $i$ and the average fitness in the population, respectively.
The dynamics given by Eq.~\eqref{cp4m} has
$n$ trivial equilibria corresponding to the consensus, i.e., the monopoly of a single state, and an interior equilibrium given by
\begin{equation}
x_i^*=\frac{s_i^{\frac{1}{1-a}}}{\sum_{\ell=1}^n s_{\ell}^{\frac{1}{1-a}}}\quad (1\le i\le n).
\label{coexistence}
\end{equation}

$V(\bm{x})\equiv-\left<sx^{a-1}\right>$, where $\bm x = (x_1, \ldots, x_n)$,
is a Lyapunov function of the dynamics given by Eq.~\eqref{cp4m} because
\begin{eqnarray}
\frac{dV(\bm{x})}{dt}&=&
-\sum_{i=1}^n s_iax_i^{a-1}\frac{dx_i}{dt} \nonumber\\
&=& -a\sum_{i=1}^n s_i x_i^a\left( s_ix_i^{a-1}-\left< sx^{a-1}\right>\right) \nonumber\\
&=& -a\left(\left<(sx^{a-1})^2\right>-\left<sx^{a-1}\right>^2\right)
\le 0.
\label{dV/dt}
\end{eqnarray}
$V$ has a unique global extremum at
$\bm x^*$, which is minimum for $a<1$ and maximum for $a>1$ (Appendix~\ref{appendix:unimodal}). Therefore, the coexistence equilibrium given by
Eq.~\eqref{coexistence} is globally stable for $a<1$ and unstable for $a>1$.

Equation~\eqref{cp4m} also admits a unique equilibrium for each
subset of the $n$ states. When $n=3$, for example, the equilibrium
in which states 1 and 2, but not 3, coexist is given by
Eq.~\eqref{coexistence} for $i=1$ and 2, with the denominator replaced
by $s_1^{\frac{1}{1-a}} + s_2^{\frac{1}{1-a}}$, and $x_3=0$. In general,
there are $\left(n\atop n^{\prime} \right)$ equilibria
containing $n^{\prime}$ states.  If $2\le n^{\prime}\le
n-1$, these equilibria are unstable.
For $a<1$, the instability immediately follows from the fact that $\bm x^*$ is the unique global minimum of $V(\bm x)$. For $a>1$, any equilibrium containing $n^{\prime}$ ($2\le n^{\prime}\le n-1$) states is unstable because it realizes the global maximum of the same Lyapunov function restricted to the simplex spanned by the $n^{\prime}$ states.

  When $a=1$, we obtain $V(\bm{x})\equiv-\left<s\right> =
  \sum_{i=1}^n s_i x_i$. Therefore, if $s_i>s_j$ ($j\neq i$), the
  consensus of state $i$ is eventually reached. If
  $s_i=s_{i^{\prime}}>s_j$ ($j\neq i, i^{\prime}$), for example, the
  $n-2$ states corresponding to $s_j$ ($j\neq i, i^{\prime}$)
are eventually eliminated. The
  dynamics then stops such that states $i$ and $i^{\prime}$ coexist. If all the
  three $s_i$ values are equal, any population is neutrally stable.

\begin{figure}[tbph]
\begin{center}
\includegraphics[width=5cm]{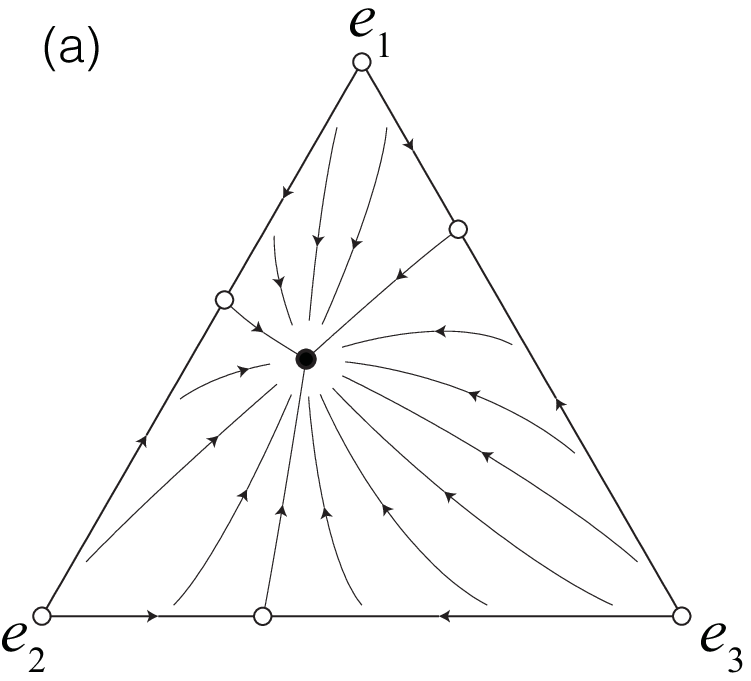}
\hspace{3mm}
\includegraphics[width=5cm]{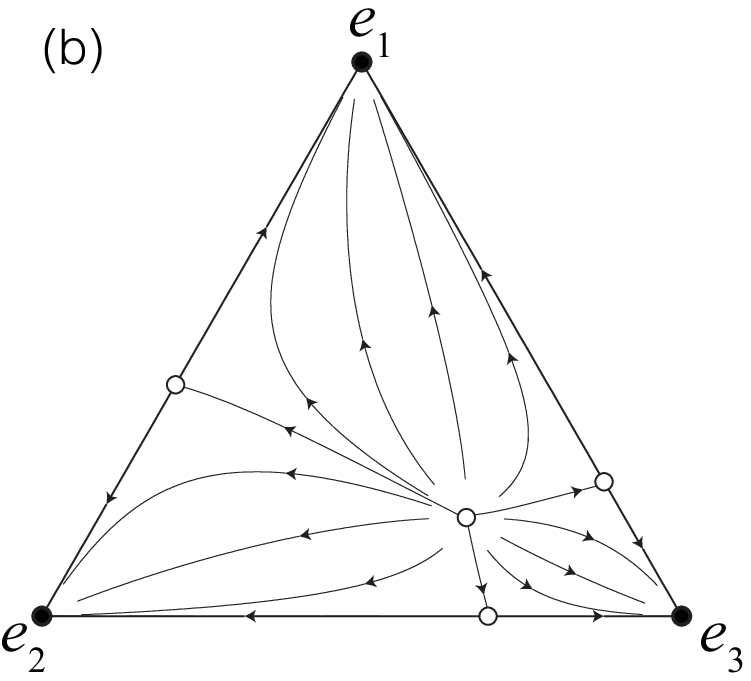}
\end{center}
\caption{Dynamics of the extended AS model when the majority preference is present and the minority aversion is absent.
We set $n=3$, $\beta=a$, $s_1=0.40$, $s_2=0.35$, and $s_3=0.25$.
(a) $a=0.5$ and (b) $a=1.4$. Solid and open circles represent stable and unstable equilibria, respectively.}
\label{flow1}
\end{figure}

Figure~\ref{flow1} represents the dynamics in the two regimes
with $n=3$, which we obtained by numerically integrating
Eq.~\eqref{cp4m}. For $a<1$, a trajectory starting from anywhere
in the interior of
the phase space, i.e., $\bm x$ that satisfies
$x_1+x_2+x_3=1$, $x_1$, $x_2$, $x_3> 0$, asymptotically approaches
the coexistence equilibrium (Fig.~\ref{flow1}a). 
It should be noted that a point in the triangle
in Fig.~\ref{flow1}a corresponds to
a configuration of the population, i.e., $\bm x$.
For example, 
corner $e_i$ ($i=1$, 2, or 3) represents the consensus (i.e., $x_i=1$ and $x_j=0$ ($j\neq i$)), and
the normalized Euclidean distance from the point to the edge $e_2$--$e_3$ of the triangle 
 is equal to the $x_1$ value. 
For $a>1$, a trajectory starting from the interior of the triangle
converges to one of the $n$ consensus equilibria,
depending on the initial condition (Fig. \ref{flow1}b).

\begin{figure}[tbph]
\begin{center}
\includegraphics[height=5.5cm]{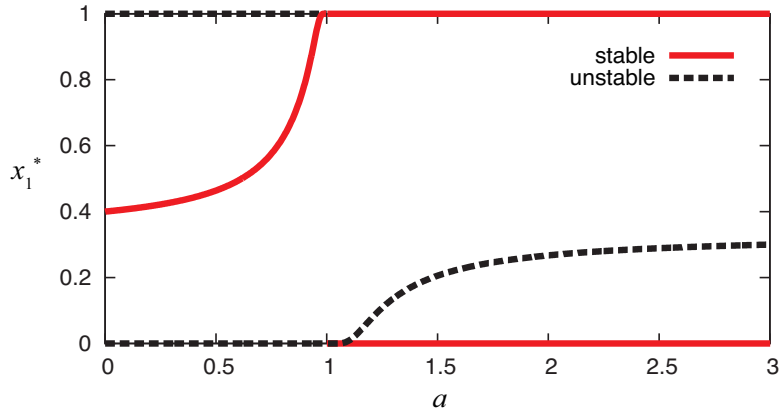}
\end{center}
\caption{Bifurcation diagram for the extended AS model when the majority preference is present and the minority aversion is absent.
The parameter values are equal to those used in Fig.~\ref{flow1} except that we vary $a$.}
\label{fixed}
\end{figure}

In Fig.~\ref{fixed},
a bifurcation diagram in which we plot
$x_1^*$ against $a$ is shown for $s_1=0.40$, $s_2=0.35$, and $s_3=0.25$.
As $a$ approaches unity from
below, the stable coexistence equilibrium approaches the unstable
consensus equilibrium corresponding to the largest $s_i$ value ($e_1$
in Fig.~\ref{flow1}).  At $a=1$, the two equilibria
collide, and an unstable coexistence equilibrium
simultaneously bifurcates from the consensus equilibrium corresponding to the smallest
$s_i$ value ($e_3$ in Fig.~\ref{flow1}).

\subsection{Case of minority aversion (i.e., $\beta=0$)}
\label{sub:minority aversion}
In this section, we set $\beta=0$ to
analyze the case in which the majority preference is absent and
the minority aversion is present.
Substitution of
Eq.~\eqref{P} in Eq.~\eqref{dx/dt} yields
\begin{eqnarray}
\frac{dx_i}{dt}&=&s_i\sum_{j=1}^n x_j(1-x_j)^a-(1-x_i)^ax_i \nonumber\\
&=&s_i\left<(1-x)^a\right>-(1-x_i)^ax_i.
\label{ca2m}
\end{eqnarray}
In contrast to the case of the majority preference (Sect.~\ref{sub:majority preference}), the simplex spanned by $n^{\prime}$ ($2\le n^{\prime}\le n-1$) states
is not invariant under the dynamics given by
Eq.~\eqref{ca2m}. Therefore, a state that once gets extinct may reappear.
In this section, we numerically analyze Eq.~\eqref{ca2m} for $n=3$ and $n=4$.
For general $n$, we analytically examine the special case in which $s_i$ is independent of $i$ in Sect.~\ref{symmetric}.

\begin{figure}[tbph]
\begin{center}
\includegraphics[width=5cm]{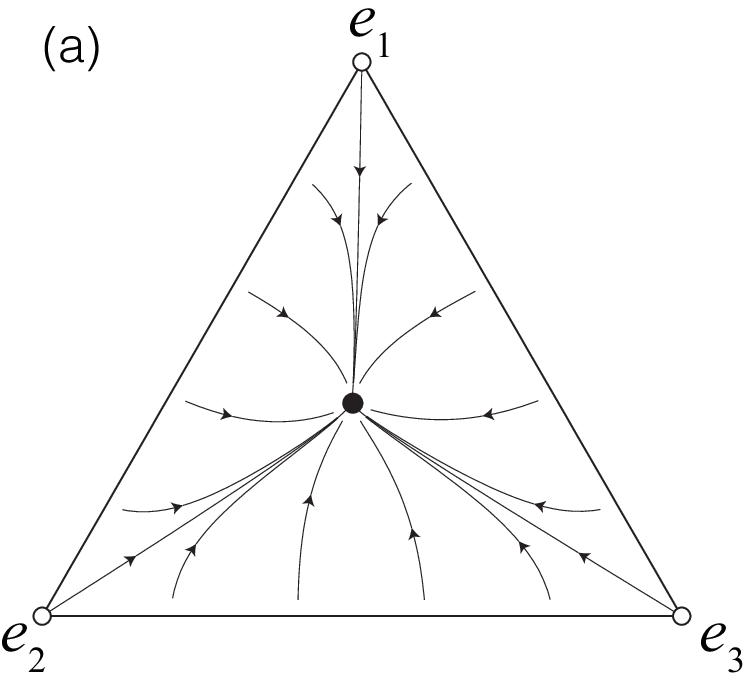}
\hspace{3mm}
\includegraphics[width=5cm]{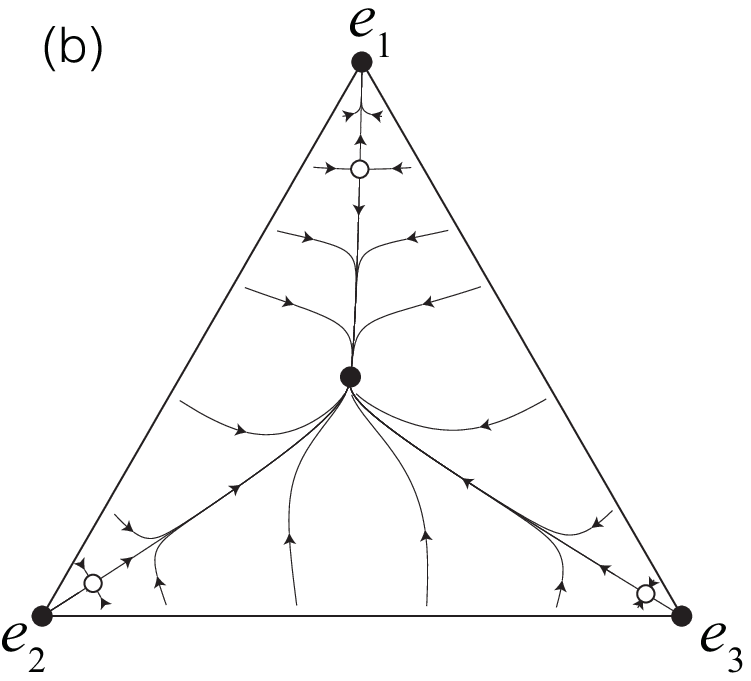}
\hspace{3mm}
\includegraphics[width=5cm]{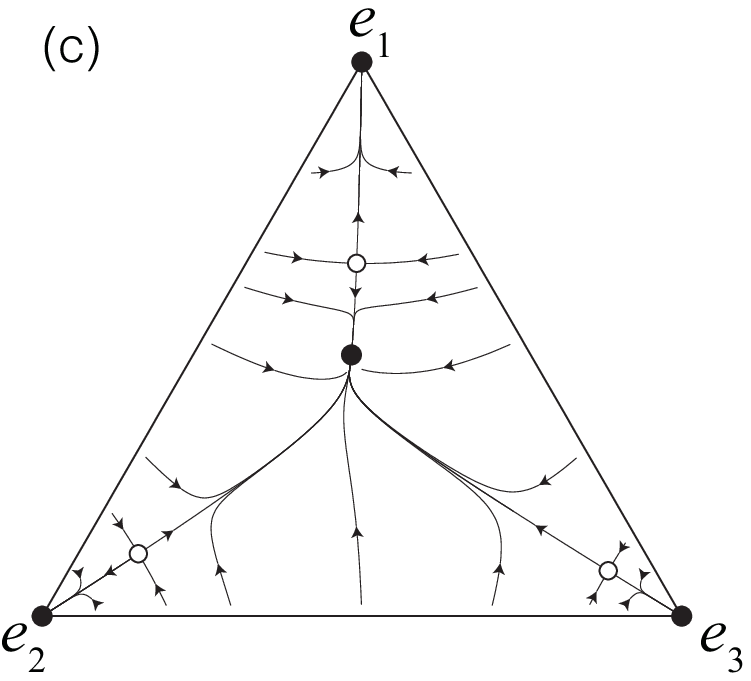}
\hspace{3mm}
\includegraphics[width=5cm]{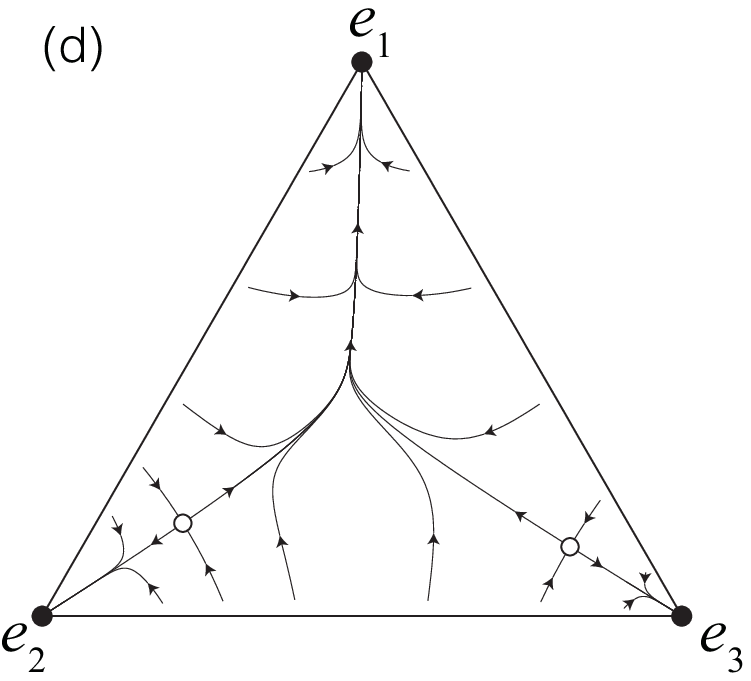}
\hspace{3mm}
\includegraphics[width=5cm]{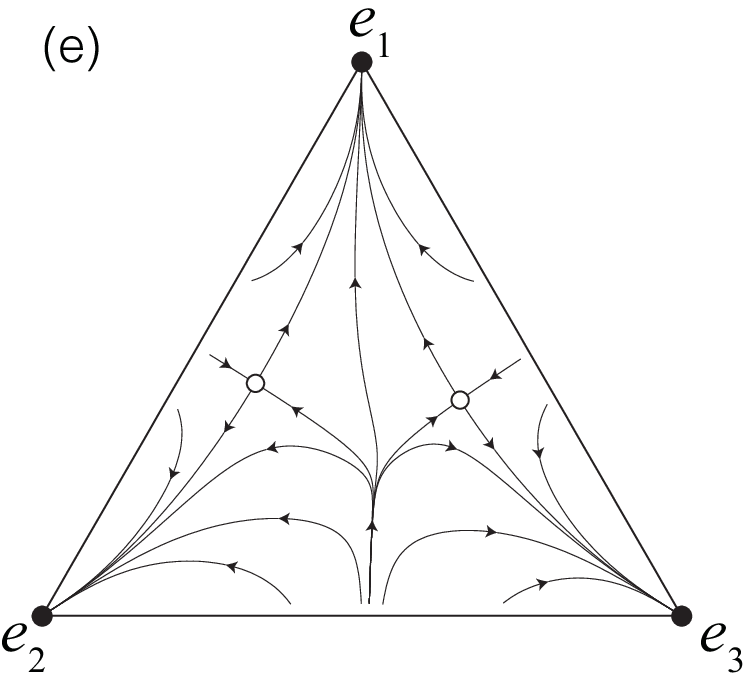}
\hspace{3mm}
\includegraphics[width=5cm]{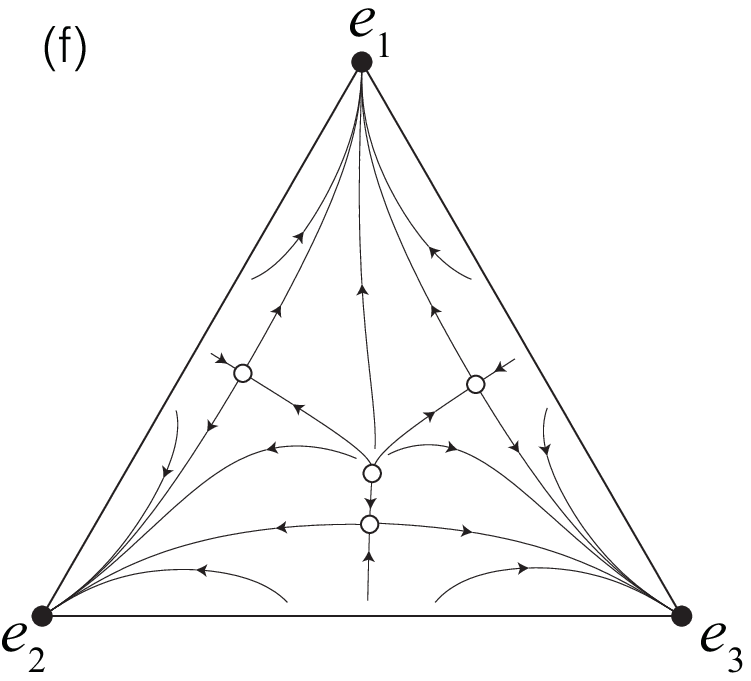}
\hspace{3mm}
\includegraphics[width=5cm]{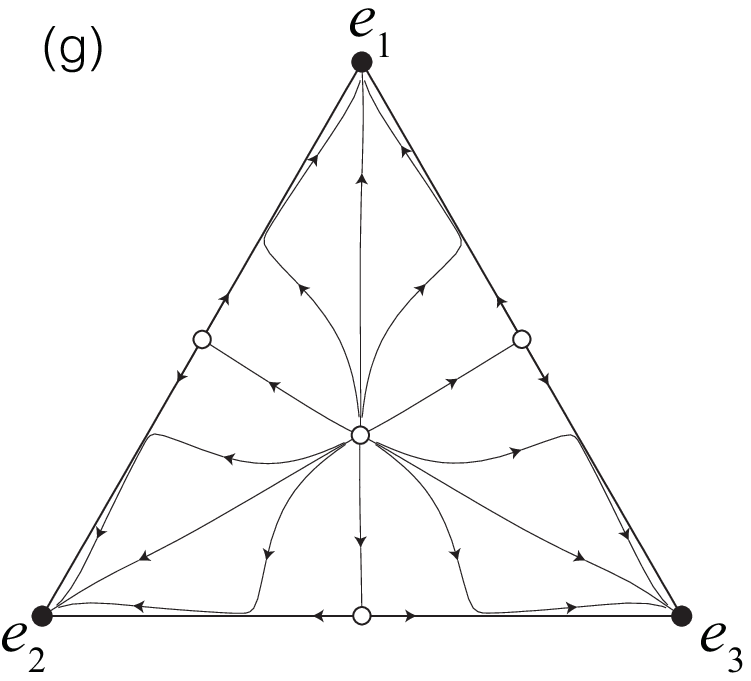}
\end{center}
\caption{Dynamics of the extended AS model when the majority preference is absent and the minority aversion is present. We set
$n=3$, $\beta=0$, $s_1=0.36$, $s_2=0.33$, and
  $s_3=0.31$. (a) $a=0.9$,
  (b) $a=1.3$, (c) $a=1.4$, (d) $a=1.5$, (e) $a=2.6$, (f) $a=2.9$, and (g)
$a=10.0$.}
\label{flow2}
\end{figure}

For $n=3$, the dynamics for various values of $a$
is shown in Fig.~\ref{flow2}. We set $s_1=0.36$, $s_2=0.33$, and $s_3=0.31$.
When $a$ is small $(a<1)$, there is a unique globally stable coexistence
equilibrium in the interior (Fig.~\ref{flow2}a).
The three consensus equilibria $e_1$, $e_2$, and $e_3$ are unstable.
At $a=1$, $e_1$, $e_2$, and $e_3$ change the stability such that they are stable beyond $a=1$ (Appendix~\ref{appendix:consensus}).
Simultaneously, a saddle point bifurcates from each consensus equilibrium. The bifurcation occurs simultaneously for the three equilibria at $a=1$ irrespective of the values of $s_1$, $s_2$, and $s_3$.
Slightly beyond $a=1$, the three consensus equilibria and the interior
coexistence equilibrium are multistable
(see Fig.~\ref{flow2}b for the results
at $a=1.3$). As
$a$ increases, the attractive basin of the coexistence
equilibrium becomes small, and
that of each consensus equilibrium becomes large (see Fig.~\ref{flow2}c
for the results at $a=1.4$).  At $a= a_{{\rm c}1}\approx 1.43$, the
coexistence equilibrium that is stable for $a<a_{\rm c1}$ 
and the unstable interior
equilibrium that bifurcates from $e_i$ at $a=1$, where $i$ corresponds to the largest $s_i$ value ($i=1$ in the present example),
collide. This is a
saddle-node bifurcation.

\begin{figure}[tbph]
\begin{center}
\includegraphics[height=5cm]{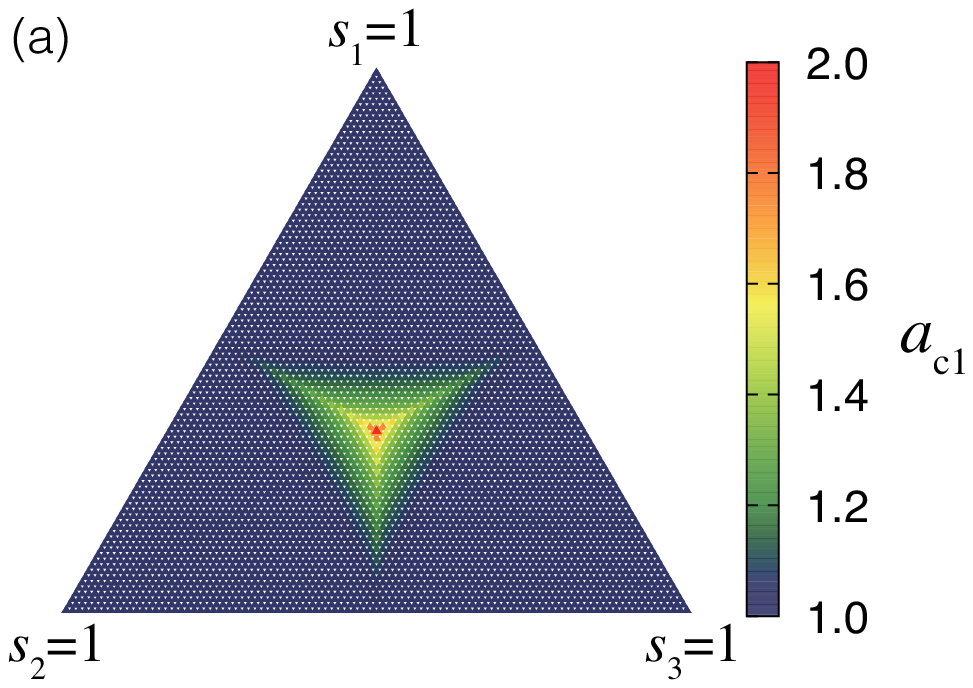}
\includegraphics[height=5cm]{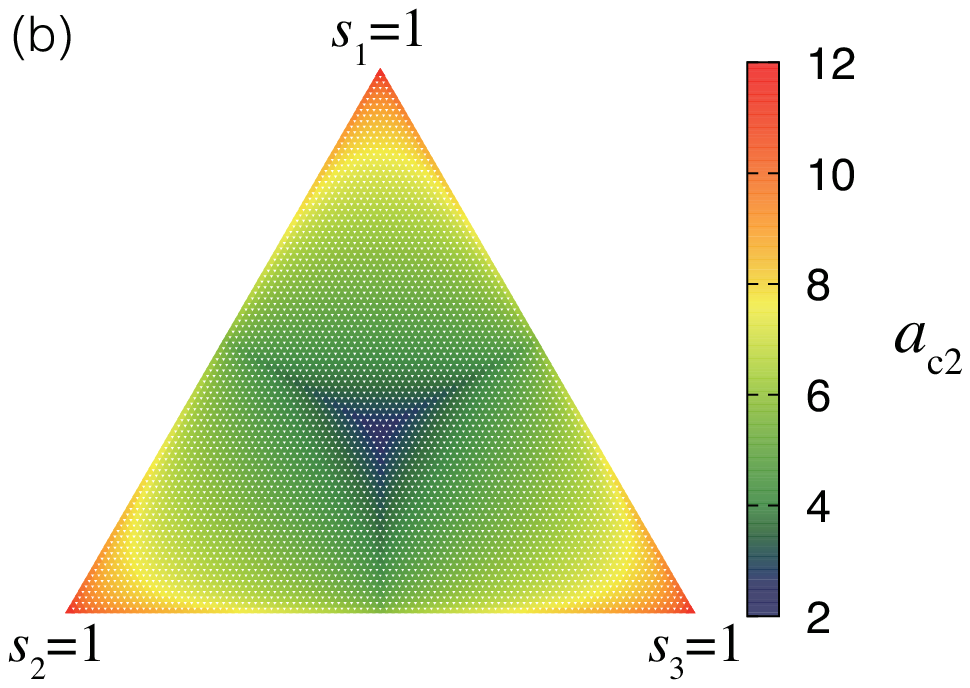}
\end{center}
\caption{Dependence of (a) $a_{\rm c1}$ and (b) $a_{\rm c2}$
on $s_1$, $s_2$, and $s_3$ when the majority preference is absent and the minority aversion is present. A point in the triangle corresponds to
a triplet ($s_1$, $s_2$, $s_3$), where $s_1+s_2+s_3=1$ and $s_i\ge 0$ ($1\le i\le 3$).}
\label{phase}
\end{figure}

Numerically obtained
$a_{\rm c1}$ values are shown in Fig.~\ref{phase}a
for different values of $s_1$,
$s_2$, and $s_3$. A point in the triangle in the figure specifies
the values of $s_1$, $s_2$, and $s_3$ under the constraint 
$s_1+s_2+s_3 = 1$, $s_i> 0$ ($1\le i\le 3$).
It seems that $a_{\rm c1}$ is the largest
when $s_i=1/3$ ($1\le i\le 3$).
Figure~\ref{phase} suggests that heterogeneity in $s_i$ makes
$a_{\rm c1}$ smaller and hence makes the stable coexistence of the three states difficult. When $s_i\approx 1$ and $s_j\approx 0$ ($j\neq i$), we obtain $a_{\rm c1}\approx 1$.

When $a$ is slightly larger than $a_{{\rm c}1}$, there are two saddle points in the interior. In this situation,
one of the three consensus equilibria, which depends on the initial condition, is eventually reached (Fig.~\ref{flow2}d). 
However, the manner with which the triangular phase space
is divided into the three attractive basins is qualitatively different from that in the case of
the majority preference (Fig.~\ref{flow1}b). In particular, in the present case of the minority aversion, even if $x_1$ is initially equal to 0, the consensus of state 1 (i.e., $e_1$) can be reached. This behavior never occurs in the case of the majority preference and less likely for a larger $a$ value
in the case of the minority aversion (Fig.~\ref{flow2}e).

\begin{figure}[tbph]
\begin{center}
\includegraphics[height=5cm]{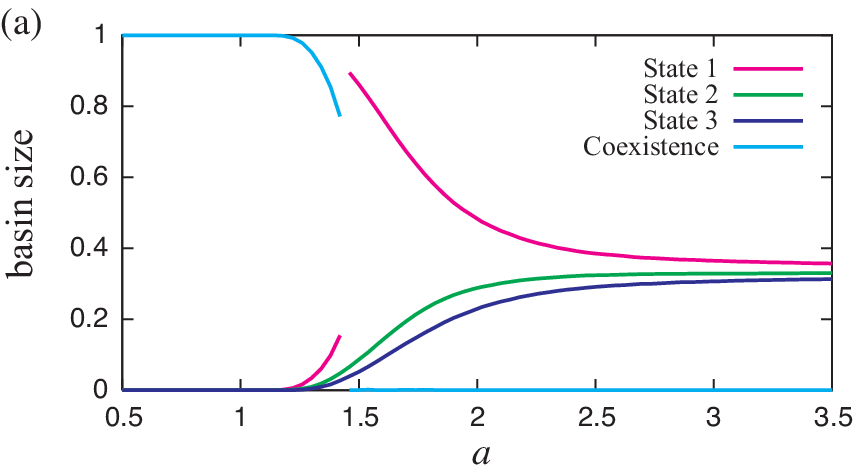}
\includegraphics[height=5cm]{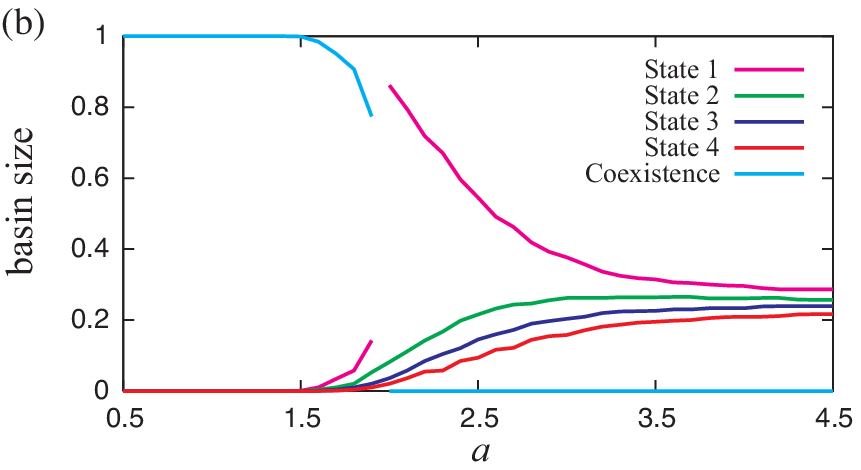}
\end{center}
\caption{Sizes of the attractive basins for
different equilibria
when the majority preference is absent and the minority aversion is present.
The lines with legend ``State $i$'' represent the basin size for the consensus equilibrium of state $i$. The lines with legend ``Coexistence'' represent the basin size for the coexistence equilibrium.
(a) $n=3$, $\beta=0$, $s_1=0.36$,
$s_2=0.33$, and $s_3=0.31$. (b) $n=4$, $\beta=0$, $s_1=0.28$,
$s_2=0.26$, $s_3=0.24$, and $s_4=0.22$. We obtain $a_{{\rm c}1}\approx
1.43$ and $a_{{\rm c}2}\approx 2.81$ in (a) and $a_{{\rm c}1}\approx
1.91$ and $a_{{\rm c}2}\approx 3.29$ in (b). We calculate the sizes of the attractive basins as follows. First, we take the initial condition ($x_1$, $x_2$, $x_3$) $=$
($0.01i$, $0.01j$, $0.01k$), where $i,j,k\ge 1$, $i+j+k=100$, for $n=3$ and
($x_1$, $x_2$, $x_3$, $x_4$) = ($0.05i$, $0.05j$, $0.05k$, $0.05\ell$), where $i,j,k,\ell\ge 1$,
$i+j+k+\ell=20$, for $n=4$. Second, we run the dynamics starting from each initial condition until the trajectory converges. Third, we count fraction of the initial conditions that converge to each stable equilibrium.}
\label{basin}
\end{figure}

The sizes of the attractive basins of the different equilibria
are plotted against $a$ in Fig.~\ref{basin}a.
Up to our numerical efforts with various initial conditions,
we did not find limit cycles.
A discrete jump in the basin size of the coexistence equilibrium
is observed at $a_{\rm c1}\approx$ 1.43, reminiscent of the saddle-node
bifurcation. Interestingly, the attractive basin of the consensus equilibrium $e_1$ is the largest just beyond $a_{\rm c1}$.

As $a$ increases further, the second saddle-node bifurcation occurs at $a=a_{\rm c2}\approx2.81$, where an unstable node and a saddle point coappear (Fig.~\ref{flow2}f). Logically, the sizes of the attractive basins could be discontinuous at $a=a_{\rm c2}$ because some initial conditions with small $x_1$ might be attracted to $e_1$ when $a$ is slightly smaller than $a_{\rm c2}$ and to $e_2$ or $e_3$ when $a$ is slightly larger than $a_{\rm c2}$. However, up to our numerical efforts, we did not observe the discontinuity, as implied by
Fig.~\ref{basin}a.

Numerically obtained
$a_{\rm c2}$ values are shown in Fig.~\ref{phase}b
for different values of $s_1$, $s_2$, and $s_3$.
Heterogeneity in $s_i$ makes $a_{\rm c2}$ larger.
In addition, $a_{\rm c2}$ is equal to $a_{\rm c1}$ when
$s_1=s_2=s_3=1/3$. In this symmetric case, the three saddle
points simultaneously collide with the stable star node at
$a=1$. Beyond $a=1$, the equilibrium that is the stable star node when $a<1$ loses
its stability to become an unstable star node.
The three saddle points move away from the unstable
star node as $a$ increases. This transition can be interpreted as
three simultaneously occurring transcritical bifurcations.

The unstable node that emerges at $a=a_{\rm c2}$ approaches $x_i^*=1/3$ ($1\le i\le 3$)
in the limit $a\rightarrow\infty$, as shown in 
Appendix~\ref{appendix:limit}.  The three saddle points
approach $(x_1, x_2, x_3)=(1/2, 1/2, 0), (1/2,
0, 1/2)$, and $(0, 1/2, 1/2)$, as shown in
Fig.~\ref{flow2}g. This is a trivial consequence of the
proof given in Appendix~\ref{appendix:limit}.
Therefore, the heterogeneity in $s_i$ does not play the role in the
limit $a\to\infty$ such that the phase space is symmetrically divided into
the three attractive basins corresponding to $e_1$, $e_2$, and $e_3$.

For $n=4$, the relationship between $a$ and the
sizes of the attractive basins of the different equilibria
is shown in Fig.~\ref{basin}b.
The results are qualitatively the same as those for $n=3$
(Fig.~\ref{basin}a).

\subsection{Symmetric case}\label{symmetric}

In the previous sections,
we separately 
considered the effect of the majority preference 
(Sect.~\ref{sub:majority preference})
and
the minority aversion (Sect.~\ref{sub:minority aversion}). In this section, we
examine the extended AS model when both effects can be combined.
To gain analytical insight into the model,
we focus on the symmetric case
$s_i=s$ ($1\le i\le n$). Although normalization $\sum_{i=1}^n s_{i}=1$ leads to $s_i=1/n$, we set $s=1$ in this section to simplify the notation; $s$ just controls the time scale of the dynamics.
Then, Eqs.~\eqref{dx/dt} and \eqref{P} are reduced to 
\begin{equation}
\frac{dx_i}{dt}=x_i^\beta\sum_{j=1, j\neq i}^n x_j(1-x_j)^{a-\beta}-x_i(1-x_i)^{a-\beta}\sum_{j=1, j\neq i}^n x_j^\beta.
\label{ndx/dt}
\end{equation}

Equation~\eqref{ndx/dt} implies that, regardless of the parameter values,
there exist $n$ trivial
consensus equilibria and symmetric coexistence equilibria of $n^{\prime}$
($2\le n^{\prime}\le n$) states given by
$x_i^*=1/n^{\prime}$, where $i$ varies over the $n^{\prime}$ surviving states arbitrarily selected from the $n$ states.

Owing to the conservation law $\sum_{i=1}^nx_i=1$, the dynamics are 
($n-1$)-dimensional. The eigenvalues of the 
Jacobian matrix of the dynamics at the coexistence equilibrium containing the
$n$ states
are ($n-1$)-fold and given by
$\left(\tfrac{1}{n}\right)^\beta\left(1-\tfrac{1}{n}\right)^{a-\beta-1}\left[(n-2)\beta+a-n+1\right]$, as shown in Appendix~\ref{appendix:coexistence}. Therefore,
the coexistence equilibrium is stable if and only if
\begin{equation}
(n-2)\beta+a-n+1<0.
\label{coexist}
\end{equation}
Similarly, we show in Appendix \ref{appendix:consensus}
that the consensus equilibria are stable if and only if
\begin{equation}
a>1.
\label{consensus}
\end{equation}
Coexistence equilibria of $n^{\prime}$ ($2\le n^{\prime}\le n-1$) states
are always unstable (Appendix \ref{appendix:coexistence}).

\begin{figure}[tbph]
\begin{center}
\includegraphics[width=8cm]{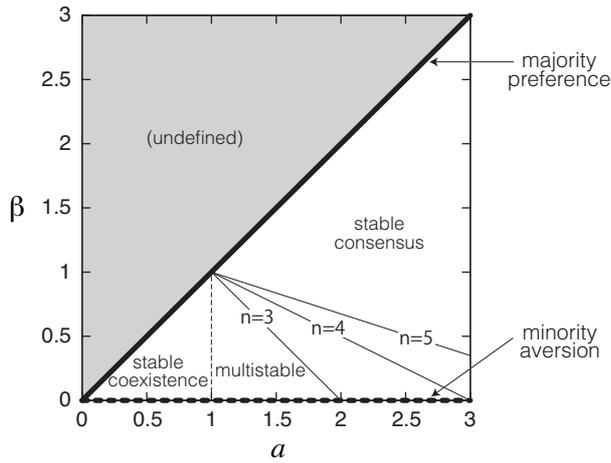}
\end{center}
\caption{Phase diagram of the extended AS model when all the $s_i$ values are equal. The thin solid lines and the thin dashed line are given by
Eqs.~\eqref{coexist}
  and \eqref{consensus}, respectively.}
\label{para_space}
\end{figure}

Figure~\ref{para_space} is the phase diagram of the model in which
the stable equilibria for given parameter values
are indicated. The thin solid and dashed lines separating two phases
are given by Eqs.~\eqref{coexist} and \eqref{consensus}, respectively.
A multistable parameter region exists when $n\ge 3$; Eq.~\eqref{coexist} is reduced to $a<1$ when $n=2$.
When $n\ge 3$, the multistablity occurs except in the case of the pure majority preference (i.e., $\beta=a$). The multistable parameter region enlarges as $n$ increases.

\section{Discussion}\label{discussion}

We analyzed an extended AS model with $n$ states.
We showed that the introduction of
the minority aversion as compared to the majority preference changes
the behavior of the model with $n\ge 3$ in two main aspects.
First, different states stably coexist up to a larger $a$ value with the minority aversion than with the majority preference. Nevertheless, it should be noted that 
$a$ is the exponent associated with different quantities in the two cases (Eq.~\eqref{P}). Second,
the multistability of the consensus
equilibria and the coexistence equilibrium is 
facilitated by the minority aversion and opposed by the majority preference.
We verified that the main results also hold true in the case of more general transition rates than Eq.~\eqref{P}, expressed as $P_{ji}=s_i^\gamma (1-s_j)^{1-\gamma}x_i^\beta(1-x_j)^{a-\beta}$
 (Appendix~\ref{sec:with gamma}).

Volovik and colleagues examined mean-field dynamics of
a three-state opinion formation model
with minority aversion
\cite{Volovik2009EPL}. Coexistence of at least
two states occurs in their model even if a random choice term, equivalent to diffusive coupling, which the authors assumed, is turned off. This is because  
only the most minor state decreases in the number of individuals
and the other two major states are equally strong in
their model.
In our model, the most major and second major states have different strengths in attracting individuals, and the coexistence equilibrium is stable
only for small $a$.

Fu and Wang considered the combined effects of the majority preference and minority avoidance on coevolution of opinions and network structure
\cite{Fu}. They assumed that the majority preference is used for collective opinion formation and the minority avoidance guides network formation.
They showed that segregated groups, each composed of individuals with the same opinion, evolve when the minority avoidance is dominantly used.
The coexistence of multiple states owing to the segregation has also been shown in spatially extended AS models \cite{Patriarca1,Patriarca2}.
We showed that coexistence of different states is facilitated by the minority aversion, not by the majority preference, and that stable coexistence 
occurs without segregation, or other spatial or network mechanisms.

Nowak and colleagues analyzed a replicator--mutator equation as a
model of language evolution \cite{Nowak}. They showed that coexistence of different grammars (i.e., states in our terminology) and a consensus-like configuration in which one grammar dominates the others are multistable when a learning parameter takes an intermediate value. If the learning is
accurate, the consensus-like configuration becomes monostable. Our model and theirs differ in at least two aspects.
First, the control parameter in our model is
$a$, that is, the strength of the sum of the majority preference and minority aversion. Second, the stable coexistence in our model requires
some minority aversion, whereas their model
takes into account the majority preference but not the minority aversion.

\begin{acknowledgements}
We thank Kiyohito Nagano, Hisashi Ohtsuki, and Gouhei Tanaka for valuable discussions.
This research is supported by the Aihara Innovative Mathematical
Modelling Project, the Japan Society for the Promotion of Science
(JSPS) through the ``Funding Program for World-Leading Innovative R\&D
on Science and Technology (FIRST Program),'' initiated by the Council
for Science and Technology Policy (CSTP).
We also acknowledge financial support provided through Grants-in-Aid for Scientific Research (No. 23681033).
\end{acknowledgements}

\appendix

\section{Unimodality of the Lyapunov function}
\label{appendix:unimodal}

We denote the Hessian of $V(\bm{x})$ by $H(V)$,
which is an $(n-1)\times (n-1)$ matrix owing to
the conservation law $\sum_{i=1}^n x_i=1$.
We obtain
\begin{eqnarray}
H_{ij}(V)&=&\frac{\partial^2V(\bm{x}) }{\partial x_i\partial x_j}\nonumber\\
&=&
 \begin{cases}
 a(1-a)\left(s_ix_i^{a-2}+s_nx_n^{a-2}\right)&(1\le i=j\le n-1),\\
 a(1-a)s_nx_n^{a-2}&(1\le i\neq j \le n-1),
 \end{cases}
\end{eqnarray}
where $x_n=1-\sum_{i=1}^{n-1}x_i$ and $s_n=1-\sum_{i=1}^{n-1}s_i$.
For any $\bm x$, $H(V)$ 
is positive (negative) definite for $a<1$ $(a>1)$ because
\begin{eqnarray}
\bm{y}^\mathrm{T}H(V)\bm{y}&=&\sum_{i=1}^{n-1}a(1-a)s_ix_i^{a-2}y_i^2+\sum_{i=1}^{n-1}\sum_{j=1}^{n-1}a(1-a)s_nx_n^{a-2}y_iy_j\nonumber\\
&=&a(1-a)\left[\sum_{i=1}^{n-1} s_ix_i^{a-2}y_i^2+s_nx_n^{a-2}\left(\sum_{i=1}^{n-1} y_i\right)^2\right]\gtrless 0\;\;\;\;\;\left(a\lessgtr 1\right),
\end{eqnarray}
for any $(n-1)$-dimensional nonzero vector $\bm{y}$.
Therefore, $V(\bm{x})$ is strictly convex (concave) for $a<1$ ($a>1$)
and hence has a unique global extremum at $\bm x^*$, which realizes
 the minimum (maximum) of $V(\bm x)$.

\section{Eigenvalues of the Jacobian matrix in the consensus equilibria}
\label{appendix:consensus}

On the basis of Eqs.~\eqref{dx/dt} and \eqref{P},
the $(n-1)\times (n-1)$ Jacobian matrix
$(J_{ij})=(\partial \dot{x_i}/\partial x_j)$ ($1\le i, j\le n-1$)
is represented by
\begin{equation}
J_{ii}=\lim_{\substack{x_i\to x_i^*\\x_k=x_k^*\;\;(k\neq i)}}\left(\sum_{j=1, j\neq i}^n k_{ji}-l_{in}-\sum_{j=1, j\neq i}^n l_{ji}+k_{in}\right)
\label{eq:J_ii}
\end{equation}
and
\begin{equation}
J_{ij}=\lim_{\substack{x_j\to x_j^*\\x_k=x_k^*\;\;(k\neq j)}}\left(l_{ij}-l_{in}-k_{ij}+k_{in}\right)\quad (i\neq j),
\label{eq:J_ij}
\end{equation}
where
\begin{eqnarray}
k_{ji}&\equiv&\beta s_ix_i^{\beta-1}x_j(1-x_j)^{a-\beta},\label{k}\\
l_{ji}&\equiv&s_j(1-x_i)^{a-\beta-1}[1-(a-\beta+1) x_i]x_j^\beta,\label{l}
\end{eqnarray}
and
\begin{equation}
x_n= 1-\sum_{i=1}^{n-1} x_i.
\label{eq:x_n}
\end{equation}

By exploiting symmetry,
we examine the Jacobian matrix at the consensus equilibrium given by $x_i=0$ ($1\le
i\le n-1$) and $x_n=1$. 
In this case, Eqs.~\eqref{eq:J_ii} and \eqref{eq:J_ij} are reduced to
\begin{equation}
J_{ii}=as_i\lim_{x_i\to 0}x_i^{a-1}-s_n
\end{equation}
and
\begin{equation}
J_{ij}=0,
\end{equation}
respectively. Therefore, irrespective of the value of $\beta$,
the eigenvalues of the Jacobian matrix, denoted by $\lambda_i$ ($1\le i\le n-1$), are given by
\begin{eqnarray}
\lambda_i=J_{ii}=
\begin{cases}
-s_n& (a>1),\\
s_i-s_n& (a=1),\\
+\infty& (a<1).
\end{cases}
\end{eqnarray}

\section{Coexistence equilibrium when $a \to \infty$}
\label{appendix:limit}

For general $n$, Eq.~\eqref{ca2m} implies that
\begin{equation}
\frac{(1-x_i)^ax_i}{s_i}=\left<(1-x)^a\right>\quad (1\le i\le n)
\end{equation}
is satisfied at the equilibrium. In particular, we obtain
\begin{equation}
\frac{(1-x_i^*)^ax_i^*}{s_i}=\frac{(1-x_j^*)^ax_j^*}{s_j}\quad (i\neq j).
\label{eq:i and j}
\end{equation}
Equation~\eqref{eq:i and j} leads to
\begin{equation}
\lim_{a\to\infty}\left(\log(1-x_i^*)+\frac{\log x_i^*}{a}-\frac{\log s_i}{a}\right)=
\lim_{a\to\infty}\left(\log(1-x_j^*)+\frac{\log x_j^*}{a}-\frac{\log s_j}{a}\right).
\label{eq:i and j 2}
\end{equation}
Because the third term on each side of Eq.~\eqref{eq:i and j 2} is negligible,
we obtain
\begin{equation}
\lim_{a\to\infty}\frac{1-x_i^*}{1-x_j^*}\left(\frac{x_i^*}{x_j^*}\right)^{\frac{1}{a}}=1.
\label{eq:i and j 3}
\end{equation}
The coexistence implies that $x_i^*$ and $x_j^*$ do not tend to 0 as $a\to\infty$. Therefore, Eq.~\eqref{eq:i and j 3} leads to $x_i^*=x_j^*$ in the limit $a\to\infty$. Therefore, $x_i^*=1/n$ ($1\le i\le n$).

\section{Eigenvalues of the Jacobian matrix in the coexistence equilibria in the symmetric case}
\label{appendix:coexistence}
On the basis of Eq.~\eqref{ndx/dt},
the $(n-1)\times (n-1)$ Jacobian matrix
$(J_{ij})=(\partial \dot{x_i}/\partial x_j)$
is represented by Eqs.~\eqref{eq:J_ii}--\eqref{eq:x_n} with
$s_i=1$ $(1\le i\le n)$.
At the coexistence equilibrium given by $x_i^*=1/n$ ($1\le i\le n$),
Eqs.~\eqref{eq:J_ii} and \eqref{eq:J_ij} are reduced to
\begin{equation}
J_{ii}=
n(k-l)=\left(\frac{1}{n}\right)^\beta\left(1-\frac{1}{n}\right)^{a-\beta-1}\left[(n-2)\beta+a-n+1\right]
\end{equation}
and
\begin{equation}
J_{ij}=0,
\end{equation}
respectively.
Therefore, the eigenvalues of the Jacobian matrix are ($n-1$)-fold degenerate and given by
\begin{equation}
\lambda_i=J_{ii}=\left(\frac{1}{n}\right)^\beta \left(1-\frac{1}{n}\right)^{a-\beta-1}\left[(n-2)\beta+a-n+1\right]\quad (1\le i\le n-1).
\end{equation}

By exploiting symmetry, we consider the coexistence equilibrium of $n^{\prime}$ states given by $x_i^* =0$ ($1\le i\le n-n^{\prime}$) and $x_i^* =1/n^{\prime}$ ($n-n^{\prime}+1\le i\le n$). In this case, Eqs.~\eqref{eq:J_ii} and \eqref{eq:J_ij} with $s_i=1$ ($1\le i\le n$) are reduced to
\begin{eqnarray}
J_{ii}&=&
\begin{cases}
\beta\left(1-\frac{1}{n'}\right)^{a-\beta} \lim_{x_i\to 0} x_i^{\beta-1}-\left(\frac{1}{n'}\right)^{\beta-1} &(1\le i\le n-n') ,\\
\left(\frac{1}{n'}\right)^\beta \left(1-\frac{1}{n'}\right)^{a-\beta-1}[(n'-2)\beta+a-n'+1]&(n-n'+1\le i\le n-1),
\end{cases}\\
J_{ij}&=&
\begin{cases}
\left(\frac{1}{n'}\right)^{\beta+1} \left(1-\frac{1}{n'}\right)^{a-\beta-1}[(n'-2)\beta+a-n'+1]&(n-n'+1\le i\le n-1\\
\qquad+\left(\frac{1}{n'}\right)^\beta-\beta\frac{1}{n'}\left(1-\frac{1}{n'}\right)^{a-\beta}\lim_{x_j\to 0}x_j^{\beta-1} &\quad\text{and } 1\le j\le n-n'),\\
0&(\text{otherwise}).
\end{cases}
\end{eqnarray}
The eigenvalues of the Jacobian matrix are the diagonal entries because the Jacobian matrix is a triangular matrix. Therefore, we obtain
\begin{eqnarray}
\lambda_i=J_{ii}&=&
\begin{cases}
-\left(\frac{1}{n'}\right)^{\beta-1} &(\beta>1) ,\\
\left(1-\frac{1}{n'}\right)^{a-1}-1&(\beta=1),\\
+\infty&(\beta<1),
\end{cases}
\label{eq:lam1}
\end{eqnarray}
for $1\le i\le n-n'$ and
\begin{equation}
\lambda_i=J_{ii}=
\left(\frac{1}{n'}\right)^\beta \left(1-\frac{1}{n'}\right)^{a-\beta-1}[(n'-2)\beta+a-n'+1]
\label{eq:lam2}
\end{equation}
for $n-n'+1\le i\le n-1$. Equations~\eqref{eq:lam1} and \eqref{eq:lam2} imply that at least one eigenvalue is positive unless $a=\beta=1$. When $a=\beta=1$, all the eigenvalues are equal to zero.

\section{Analysis of a generalized model}\label{sec:with gamma}

Consider a variant of the extended AS model in which the transition
rate given by Eq.~\eqref{P} is replaced by
\begin{equation}
P_{ji}=s_i^\gamma (1-s_j)^{1-\gamma}x_i^\beta(1-x_j)^{a-\beta},
\label{P2}
\end{equation}
where $0\le\gamma\le 1$. Equation~\eqref{P} is reproduced with $\gamma=1$.
Here, the exponents $\gamma$ and $1-\gamma$ represent the strength of the preference for a strong state and the aversion to a weak state, respectively.

In the case of the majority preference (i.e., $\beta=a$), Eq.~\eqref{dx/dt} with the transition rates given by Eq.~\eqref{P2} has 
$n$ trivial equilibria corresponding to the consensus and an interior equilibrium given by
\begin{equation}
x_i^*=\frac{\left(s_i^{-\gamma}(1-s_i)^{1-\gamma}\right)^{\frac{1}{a-1}}}{\sum_{\ell=1}^n\left(s_{\ell}^{-\gamma}(1-s_{\ell})^{1-\gamma}\right)^{\frac{1}{a-1}}}.
\label{coexistence2}
\end{equation}
There also exists a unique equilibrium composed of 
arbitrarily chosen $n^{\prime}$ states ($2\le n^{\prime}\le n-1$).
The Lyapunov function is given by
\begin{eqnarray}
V(\bm{x})=-\frac{\left<s^\gamma x^{a-1}\right>}{\left<(1-s)^{1-\gamma}\right>^a}.
\label{Lyapunov2}
\end{eqnarray}
$V(\bm x)$ is a Lyapunov function because
\begin{eqnarray}
\frac{d\left<s^\gamma x^{a-1}\right>}{dt}&=& \sum_{i=1}^n s_i^\gamma ax_i^{a-1}\frac{dx_i}{dt} \nonumber\\
&=& a\sum_{i=1}^n s_i^\gamma x_i^{a-1}\left[ s_i^\gamma x_i^a\left<(1-s)^{1-\gamma}\right>-(1-s_i)^{1-\gamma}x_i\left< s^\gamma x^{a-1}\right>\right] \nonumber\\
&=& a\left(\left<(s^\gamma x^{a-1})^2\right>\left<(1-s)^{1-\gamma}\right>-\left<s^\gamma x^{a-1}(1-s)^{1-\gamma}\right>\left<s^\gamma x^{a-1}\right>\right)
\label{dav1}
\end{eqnarray}
and
\begin{eqnarray}
\frac{d\left<(1-s)^{1-\gamma}\right>}{dt}&=& \sum_{i=1}^n (1-s_i)^{1-\gamma} \frac{dx_i}{dt} \nonumber\\
&=& \sum_{i=1}^n (1-s_i)^{1-\gamma}\left[ s_i^\gamma x_i^a\left<(1-s)^{1-\gamma}\right>-(1-s_i)^{1-\gamma}x_i\left< s^\gamma x^{a-1}\right>\right] \nonumber\\
&=& \left<s^\gamma x^{a-1}(1-s)^{1-\gamma}\right>\left<(1-s)^{1-\gamma}\right>-\left<(1-s)^{2(1-\gamma)}\right>\left<s^\gamma x^{a-1}\right>
\label{dav2}
\end{eqnarray}
lead to
\begin{eqnarray}
\frac{dV(\bm{x})}{dt}&=&-\frac{\frac{d}{dt}\left<s^\gamma x^{a-1}\right>}{\left<(1-s)^{1-\gamma}\right>^a}+\frac{a\left<s^\gamma x^{a-1}\right>\frac{d}{dt}\left<(1-s)^{1-\gamma}\right>}{\left<(1-s)^{1-\gamma}\right>^{a+1}}\nonumber\\
&=&-\frac{a\left(\left<A^2\right>\left<B\right>^2-2\left<AB\right>\left< A\right>\left<B\right>+\left< A\right>^2\left<B^2\right>\right)}{\left<B\right>^{a+1}} \nonumber \\
&=&-\frac{a\left<\left(A\left<B\right>-\left<A\right>B\right)^2\right>}{\left<B\right>^{a+1}} \;\;\;\; \le 0,
\end{eqnarray}
where $A\equiv s^\gamma x^{a-1}$ and $B\equiv (1-s)^{1-\gamma}$.

We have no proof of the unimodality of $V(\bm x)$.
However, we numerically verified the unimodality for $n=3$.
These results are the same as those obtained in the main text for $\gamma=1$
(Sect.~\ref{sub:majority preference}).

For any $\beta$, the
Jacobian matrix $(J_{ij})=(\partial \dot{x_i}/\partial x_j)$ is given by Eqs.~\eqref{eq:J_ii} and \eqref{eq:J_ij}, where
\begin{eqnarray}
k_{ji}&\equiv&\beta s_i^\gamma(1-s_j)^{1-\gamma} x_i^{\beta-1}x_j(1-x_j)^{a-\beta},\label{k2}\\
l_{ji}&\equiv&s_i^\gamma(1-s_j)^{1-\gamma} (1-x_j)^{a-\beta-1}[1-(a-\beta+1) x_j]x_i^\beta,\label{l2}
\end{eqnarray}
and
\begin{equation}
x_n= 1-\sum_{i=1}^{n-1} x_i.
\end{equation}
In the consensus equilibrium given by $x_i=0$ ($1\le
i\le n-1$) and $x_n=1$,
Eqs.~\eqref{eq:J_ii} and \eqref{eq:J_ij} are reduced to
\begin{equation}
J_{ii}=as_i^{\gamma}(1-s_n)^{1-\gamma}\lim_{x_i\to 0}
x_i^{a-1}-s_n^{\gamma}(1-s_i)^{1-\gamma}
\end{equation}
and
\begin{equation}
J_{ij}=0,
\end{equation}
respectively.
Therefore, the eigenvalues of the Jacobian matrix are given by
\begin{eqnarray}
\lambda_i=J_{ii}=
\begin{cases}
-s_n^\gamma(1-s_i)^{1-\gamma}& (a>1),\\
s_i^\gamma(1-s_n)^{1-\gamma}-s_n^\gamma(1-s_i)^{1-\gamma} & (a=1),\\
+\infty& (a<1),
\end{cases}
\end{eqnarray}
and the consensus equilibrium is stable if $a>1$.

\begin{figure}[tbph]
\begin{center}
\includegraphics[height=5cm]{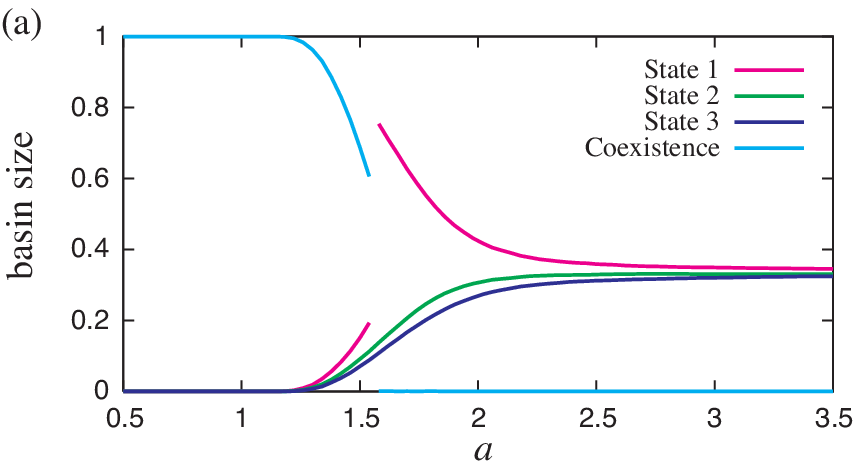}
\includegraphics[height=5cm]{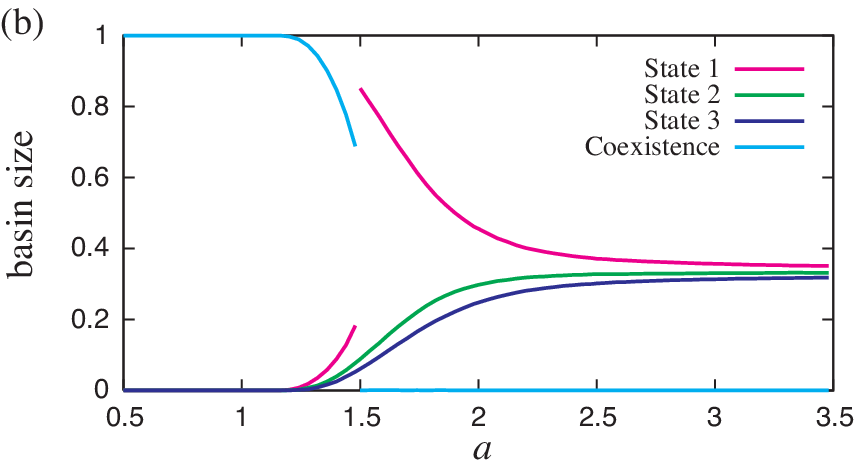}
\end{center}
\caption{Sizes of the attractive basins for
different equilibria 
when the majority preference is absent and the minority aversion is present
in a generalized model. We set (a) $\gamma=0$ and (b) $\gamma=0.5$.
We also set $n=3$, $\beta=0$, $s_1=0.36$,
$s_2=0.33$, and $s_3=0.31$, the same parameter values as those used for
Fig.~\ref{basin}a. 
We obtain $a_{{\rm c}1}\approx
1.57$ and $a_{{\rm c}2}\approx 2.56$ in (a) and $a_{{\rm c}1}\approx
1.50$ and $a_{{\rm c}2}\approx 2.69$ in (b). The procedure for calculating the basin size is described in the caption of Fig.~\ref{basin}.}
\label{fig:basin different gamma}
\end{figure}

In the case of the minority aversion (i.e., $\beta=0$), when $n=3$, the sizes of the attractive basins for different equilibria for $\gamma=0$ and
$\gamma=0.5$ are shown in 
Figs.~\ref{fig:basin different gamma}a and \ref{fig:basin different gamma}b, respectively.
The results are qualitatively the same as those for $\gamma=1$
(Fig.~\ref{basin}a).

\end{document}